\documentclass[journal]{IEEEtran}

\usepackage{amssymb}
\usepackage{amsmath}
\usepackage{graphicx}
\usepackage{setspace}
\usepackage{euscript}
\usepackage{balance}
\usepackage{cite}

\begin{document}
%
\title{Hadamard Coded Modulation for \\Visible Light Communications}

\author{{Mohammad Noshad,~\IEEEmembership{Member,~IEEE}, and Ma\"{\i}t\'{e} Brandt-Pearce,~\IEEEmembership{Senior Member,~IEEE}}
\thanks{Mohammad Noshad (noshad@vlncomm.com) is with VLNComm LLC, Charlottesville, VA 22903.}
\thanks{Ma\"{\i}t\'{e} Brandt-Pearce (mb-p@virginia.edu) is with the Charles L. Brown Department of Electrical and Computer Engineering, University of Virginia Charlottesville, VA 22904.}
\thanks{This work was supported in part by the US Department of Energy (DOE) through SBIR program under grant number DE-SC0013195.}
}

\markboth{}{Shell \MakeLowercase{\textit{et al.}}: Bare Demo of IEEEtran.cls for Journals}

\maketitle
\thispagestyle{empty}
\pagestyle{empty}

\begin{abstract}
Visible light communication (VLC) systems using the indoor lighting system to also provide downlink communications require high average optical powers in order to satisfy the illumination needs. This can cause high amplitude signals common in higher-order modulation schemes to be clipped by the peak power constraint of the light emitting diode (LED) and lead to high signal distortion.
In this paper we introduce Hadamard coded modulation (HCM) to achieve low error probabilities in LED-based VLC systems needing high average optical powers. This technique uses a fast Walsh-Hadamard transform (FWHT) to modulate the data as an alternative modulation technique to orthogonal frequency division multiplexing (OFDM). HCM achieves a better performance for high illumination levels because of its small peak to average power ratio (PAPR). The power efficiency of HCM can be improved by reducing the DC part of the transmitted signals without losing any information.  The resulting so-called DC-reduced HCM (DCR-HCM) is well suited to environments requiring dimmer lighting as it transmits signals with lower peak amplitudes compared to HCM, which are thus subject to less nonlinear distortion. Interleaving can be applied to HCM to make the resulting signals more resistant against inter-symbol interference (ISI) in dispersive VLC links.
\end{abstract}


\section{Introduction}  \label{sec: Introduction}

\IEEEPARstart{V}{ISIBLE} light communications (VLC) is an emerging technology for indoor wireless networking that can offer energy efficient Gbps streaming through the lighting system. The idea is to transmit downlink data by modulating white light emitting diodes (LED) that are already being used by energy efficient and cost effective lighting systems. The huge unregulated bandwidth available in VLC technology can relieve the traffic on radio-frequency (RF) communications. However, using LEDs as sources adds restrictions on the modulation schemes and codes that can be used. Main limitations of these LEDs are their limited peak optical power, nonlinear transfer function, and limited modulation bandwidth \cite{VLC-Ghassemlooy}. Therefore, modulation and coding schemes with high spectral efficiencies are required to provide a high data-rate connection.
In this work we propose a new modulation technique to achieve reliable and high-speed data transmission in nonlinear-LED-based VLC systems.


Orthogonal frequency-division multiplexing (OFDM) is an efficient modulation technique for high speed data communication through bandlimited channels, and is being widely used in modern  systems because of its high spectral efficiency and robustness against narrow-band interference \cite{OFDM-Book}. OFDM signals are generated by applying an inverse fast Fourier transform (IFFT) on the data stream at the transmitter and decoded using a fast Fourier transform (FFT) at the receiver.
OFDM has been adapted to work in energy efficient optical communications because of its better average-power efficiency compared to other schemes \cite{Optical-OFDM-Armstrong-13}.
Asymmetrically clipped optical OFDM (ACO-OFDM), DC-biased optical OFDM (DCO-OFDM) and asymmetrically clipped DC biased optical OFDM (ADO-OFDM) are modified forms of OFDM proposed for intensity-modulation direct-detection (IM/DD) optical communication systems \cite{ACO-OFDM-06, DCO-OFDM-06, ADO-OFDM-11}. These OFDM techniques use Hermitian symmetry to generate real signals from the data sequence, trading-off loss in the encoding rate. Due to the low rate of these techniques, high order quadratic amplitude modulation (QAM) has to be employed to achieve high spectral efficiencies, which degrades the energy efficiency.

As in the original OFDM, these optical communications techniques generate signals with large peaks, which can end up clipped by the peak optical power constraint of the optical sources. This clipping causes a distortion of the OFDM signals that becomes larger by increasing the transmitted average power. Consequently, in VLC systems where high average optical powers are required for illumination, OFDM is of limited use. This problem can be alleviated by using peak to average power ratio (PAPR) reduction techniques that trade-off complexity and energy inefficiency \cite{OFDM-PAPR-Reduction-VLC-Zabih-14}. For example, Hadamard matrices can be used as precoders in OFDM systems to decrease the PAPR \cite{OFDM-Hadamard-11, OFDM-Hadamard-12}, reduce the BER \cite{OFDM-Hadamard-BER-03} and increase the resistance of the signals against frequency selective fading \cite{OFDM-Hadamard-Fading-10}.

The challenge of supporting a wide range of dimming levels is another big drawback in the application of these modified forms of OFDM to VLC systems. There have been significant efforts to address this problem \cite{OFDM-for-VLC-Wang-14}. Reverse polarity optical-OFDM (RPO-OFDM) combines pulse width modulation (PWM) with OFDM to change the dimming level of the transmitted signals \cite{Dimming-for-OFDM-13}. Although this technique can transmit signals with high average optical power without being significantly clipped by the LED peak-power, the lowest BER it can achieve is higher than that of ACO-OFDM and DCO-OFDM.

Pulsed modulation techniques are another solution to achieve reliable high-speed communication at high optical average power levels in VLC systems \cite{PWM-PPM-VLC-Zabih-14}. Among these techniques, those that use the optical sources in their on/off mode are preferred since they avoid the nonlinear effects of the LEDs \cite{VLC-Networking13}. Although using sources only in their on/off mode limits the spectral efficiency of the system in single-LED systems \cite{EPPM12}, systems with multiple LED sources have the potential to use multilevel signaling, which can be employed to use the available bandwidth more efficiently \cite{Multilevel-EPPM12}.

In \cite{HCM-Globecom-14} we introduce a multilevel modulation technique named HCM that uses the Hadamard matrices as a modulation technique (rather than a precoder). In this technique, the data is modulated using a fast Walsh-Hadamard transform (FWHT) and the receiver uses an inverse fast Walsh-Hadamard transform (IFWHT) to decode the received signals. Therefore, it has a low complexity and can exploit the bandwidth effectively. We also propose a modification referred to as the DC-reduced HCM (DCR-HCM) technique.  Because of their low PAPR, HCM and DCR-HCM can provide high illumination levels in VLC systems without being too affected by LED-induced distortion. The technique is reminiscent of the approach in \cite{Hadamard-in-CDMA-04} to use the bipolar Hadamard transform for channel orthogonalization in code division multiple access (CDMA) systems because of their low PAPR.

In this paper we analyze these modulations and study their performance in LED-based VLC systems. We propose using sinc pulses for efficient use of the available LED bandwidth bandwidth and for fair comparison with OFDM. We show that using DCR-HCM the energy efficiency of HCM can be improved. This improvement becomes more significant by increasing the size of the FWHT. DCR-HCM is also able to achieve lower BER levels compared to HCM and OFDM due to its reduced DC level, which decreases the amplitudes of the transmitted signals and makes them less likely to be clipped by the peak-power limit of the LEDs. As in OFDM systems, a cyclic prefix is used to avoid interference between adjacent symbols in bandlimited channels, and then symbol-length interleaving is applied on the HCM signals, as was proposed in \cite{VLC-JLT-I-13}, to decrease the effect of the intra-symbol interference. This approach is shown to lower the error probability of high data-rate transmissions through VLC channels.

The rest of the paper is organized as follows. Section~\ref{sec:System Description} describes the VLC system model including the LED nonlinear transfer function and VLC channel characteristics. Section~\ref{sec:HCM} introduces the principles of HCM and presents modified forms of HCM to increase its energy efficiency and make it renitent against inter-symbol interference (ISI) in dispersive VLC channels. Numerical results are presented in Section~\ref{sec:Numerical Results} that compare the performance of the HCM and its modified forms to OFDM in VLC systems. Finally, conclusions are drawn in Section~\ref{sec:Conclusion}.

\section{Problem Description} \label{sec:System Description}
This section describes the principles of a VLC system. Models for the optical sources and VLC channel are discussed.

In this work we represent vectors with boldfaced lower-case letters, and boldfaced upper-case letters are reserved for matrices. The identity matrix is represented by $\mathbf{I}$. The notation $\mathbf{A}^{\text{T}}$ denotes the transpose of the matrix $\mathbf{A}$, and $x^*$ indicates the conjugate of the complex number $x$. The notations $\mathbf{A}-x$ and $\mathbf{a}-x$ are respectively used to show the matrix and vector that are obtained by subtracting a scalar $x$ from all elements of the matrix $\mathbf{A}$ and vector $\mathbf{a}$. We define the complement of the vector $\mathbf{a}$ as $\overline{\mathbf{a}}:=(1-\mathbf{a})$, and that of the matrix $\mathbf{A}$ as $\overline{\mathbf{A}}:=(1-\mathbf{A})$.
In this paper, $\mathbf{a}{^{(\ell)}}$ and $\mathbf{A}{^{(\ell)}}$ represent the $\ell$th right cyclic-shifts of vector $\mathbf{a}$ and matrix $\mathbf{A}$, respectively.

\subsection{VLC System Description} \label{sec:VLC Config}



According to the Illuminating Engineering Society of North America (IESNA), the standard illumination level for most indoor environments (classroom, conference-room, lecture hall, offices, etc.) is between 300 and 500 lux at 0.8 m height from the floor \cite[Table 32.I]{Illuminance-Standards}. In the daytime, a portion of the indoor illumination needs could be provided by daylight, and the lights can then be dimmed to reduce the energy consumed. The dimming level has a nonlinear relation to the average optical power \cite{VLC-Lighting-Requirements-13}, and affects the performance of the VLC system.

In order to evaluate the performance of a VLC link, it is essential to determine the optical power level that corresponds to a desired illumination level. This can be done using luminance efficiency of radiation (LER), which is defined as the luminous flux per unit optical power. Although the theoretical limit of LER for white LEDs is 260-300 lm/W \cite{LED-Efficiency-Limit-10}, commercially available white LEDs have an LER of 50-150 lm/W. One of the most efficient currently available white LEDs is Cree's XLamp XT-E white LED with an LER of 148 lm/W.  Since 1 lux = 1 lm/m$^2$, an illumination level of 500 lux from a light source with an LER of 148 lm/W corresponds to 33.8 $\mu$W average optical power on a photo-detector with an effective area of 0.1 cm$^2$.

\subsection{Nonlinearity in Optical Sources} \label{sec:LED Nonlinearity}
The optical source is a key component of any optical communication system, as it generates optical power as a function of the modulated input electrical signal and converts the information into an optical beam. Because of the structure of LEDs, the output optical power and the forward current are related by a nonlinear function. The maximum optical power in these sources is limited to a peak-power, and this can result in the clipping of large peaks in the modulated signal.
In optical communication systems using multilevel or continuous valued signaling, the nonlinearity of the optical sources introduces a distortion on the transmitted optical signal.


Predistorion is a solution to linearize the relation between the output optical power and the forward current over a range. This technique requires an accurate model for the design of the predistortion and linearization over the dynamic range of the optical source. A polynomial model is presented in \cite{Optical-OFDM-Nonlinear-LED-13} to describe the nonlinear transfer function of the optical sources, through which a predistortion function can be designed to linearize the output optical power in terms of the input current.

A problem with the predistorion technique is that the nonlinear transfer function of optical sources can change due to many factors, one of which is the temperature of the transmitter. LEDs and lasers tend to dissipate a portion of the input energy as heat, which increases the temperature of the device over time and changes the nonlinear relation between the output power and forward current. This means that one predistortion function is not able to keep the device linear over time, and dynamic feedback is needed to modify the model of the instantaneous nonlinear transfer function of the optical source \cite{OWC-IR-Book-08} and actively match the predistortion to that model in order to keep the relation between the optical power and input current linear. This makes the design of the transmitter more complex.  In this paper we assume no predistortion is employed.

As discussed in \cite{VLC-JLT-I-13}, in VLC systems that employ arrays of LEDs as sources, pulsed modulation techniques can solve the problem by using the LEDs in an on/off mode.
In these modulation techniques, multilevel signals can be generated by independently turning on and off each element of the LED-array. In this way multilevel signaling can be used without concern for the effects of the LED nonlinearity, and the optical signal level remains proportional to the intended modulating signal \footnote{Note that the LED nonlinearity could still affect the pulse-shape, an effect that is ignored in this paper.}. Based on a similar idea, quantized OFDM is proposed to utilize the full dynamic range of LEDs by using LED arrays and employing discrete power level stepping \cite{LED-Level-Quantization-Haas-13}. Below we show that HCM symbols can also be generated using an LED-array without being affected by the LED nonlinearity.

\begin{figure} [!t]
    \begin{center}
    {\includegraphics[width=3.4in]{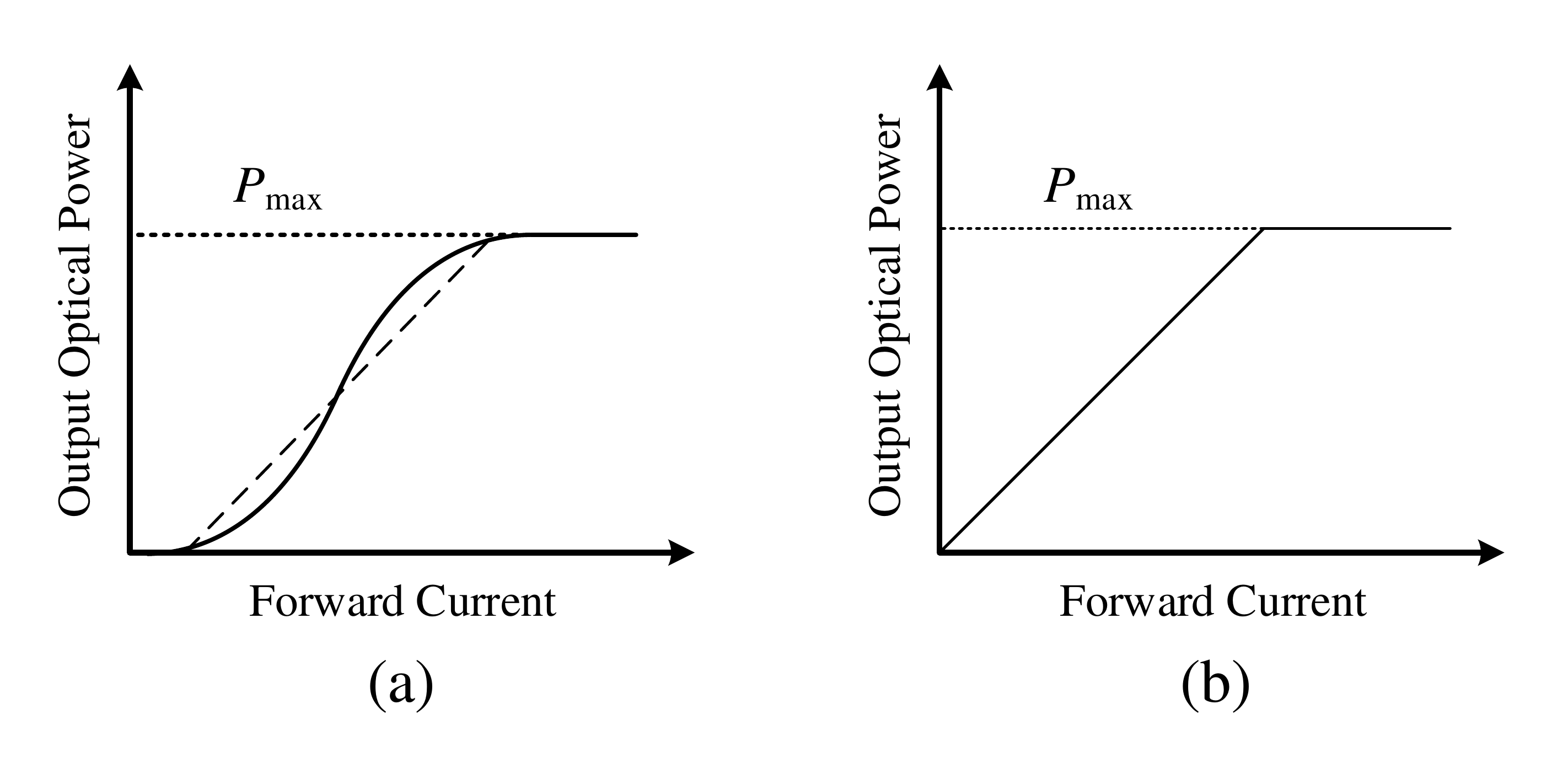}}
    \end{center}
    \vspace{-0.2in}
    \caption{(a) Nonlinear transfer function of an LED (solid) and its linearization after pre-distortion (dashed), and (b) transfer function of and ideal LED with a limited peak-power.}
    \label{Fig:LED-Transfer Function}
    \vspace{-0.25in}
    \end{figure}

In this paper we model the LED as an ideal peak-power limited source, i.e., a hard limiter, which generates a power ranging from 0 to the LED peak power, $P_{\max}$, proportional to the forward input current (Fig.~\ref{Fig:LED-Transfer Function}-(b)). Ignoring the bandwidth limit of the LED, the only distortion on the transmitted signals is assumed to be caused by clipping the transmitted signals at 0 and the peak-power of the LED. We model the clipping induced distortion by an attenuation and an additive Gaussian noise with variance \cite{Amp-Distortion-Bussgang}
    \begin{align}\label{Variance of Clipping Distortion}
        \sigma{^2_{\text{clip}}} = \int_{-\infty}^{0} x^2 f(x) dx + \int_{P_{\max}}^{\infty} \left(x-P_{\max}\right)^2 f(x) dx ,
    \end{align}
where $f(\cdot)$ is the probability distribution function (pdf) of the amplitude of the signal sent to the LED.

\subsection{VLC Channel Model} \label{sec:Channel Model}
\begin{figure} [!t]
    \begin{center}
    {\includegraphics[width=3.4in]{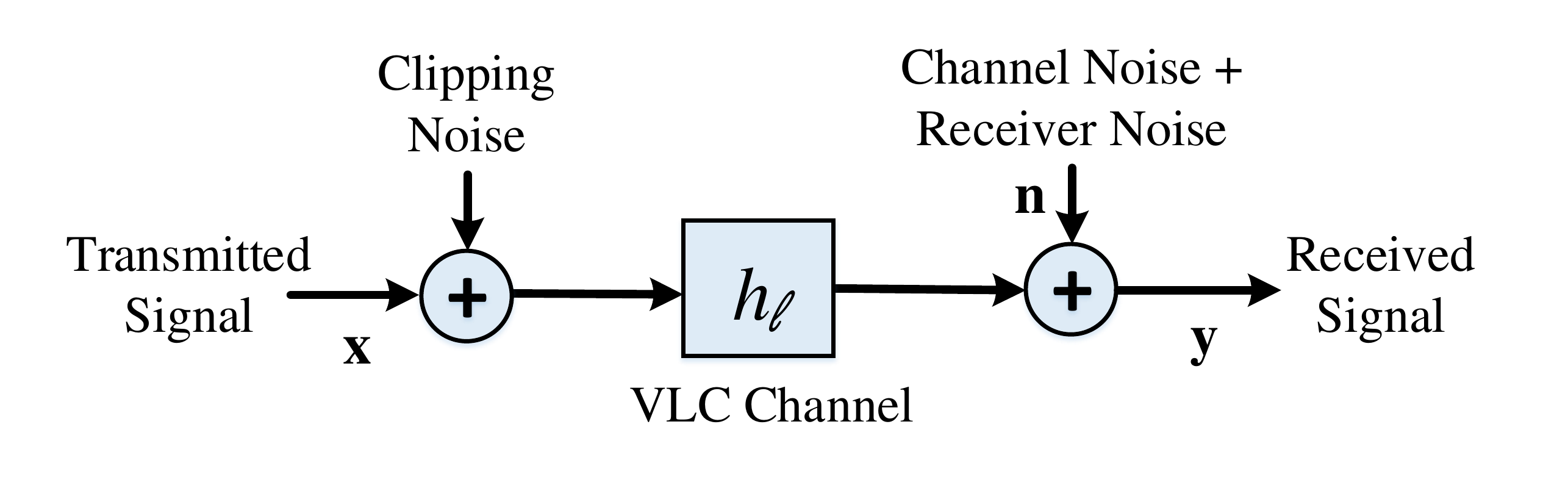}}
    \end{center}
    \vspace{-0.2in}
    \caption{VLC system front end.}
    \label{Fig:VLC-Model}
    \vspace{-0.2in}
    \end{figure}

The impulse-response of a VLC channel consists of line-of-sight (LOS) and non-line-of-sight (NLOS) parts. In VLC systems, the NLOS part of the impulse response is due to  reflections of the light from the walls and other objects and usually causes inter-symbol interference at symbol-rates  higher than 50 Msps. The results of \cite{OWC-Channel05} and \cite{OWC-Channel11} can be used to find the impulse-response of a VLC channel with a given room geometry. Given the sampling period, an equivalent discrete impulse-response of the VLC channel, $\mathbf{h} = \{h_{\ell} \}$, can be calculated from the continuous impulse-response.
For mathematical simplicity, in this work we assume the normalized discrete impulse-response of a VLC channel for which we have $\sum\limits_{\ell=-\infty}^{\infty} {h_{\ell}} = 1$, and the channel loss is ignored for notational convenience. We model the noise in the system as an additive white Gaussian noise (AWGN) source, which is a good approximation for high background light scenarios. The front-end of our VLC system model is depicted in Fig.~\ref{Fig:VLC-Model}.

\section{Hadamard Coded Modulation} \label{sec:HCM}
Hadamard coded modulation (HCM), which uses a binary Hadamard matrix to modulate the input data, is introduced in \cite{HCM-Globecom-14} as an alternative to OFDM. As described in \cite{HCM-Globecom-14}, the HCM signal $\mathbf{x}=[x_0,x_1,\cdots,x_{N-1}]^{\textmd{T}}$ is generated from the data sequence $\mathbf{u}=[0,u_1,\cdots,u_{N-1}]^{\textmd{T}}$ as
    \begin{align}\label{HCM Encoder}
        \mathbf{x} = \Bigg( \mathbf{H}{_N} \mathbf{u} + \overline{\mathbf{H}}{_N} \overline{\mathbf{u}} \Bigg).
    \end{align}
where $\mathbf{H}_N$ is the binary Hadamard matrix of order $N$ \cite{Hadamard-Book}, $\mathbf{\overline{H}}_N$ is the complement of $\mathbf{H}_N$, and the matrix $( \mathbf{H}{_N} - \overline{\mathbf{H}}{_N})$ is the bipolar Hadamard matrix.


The components of $\mathbf{u}$ are assumed to be $M$-ary pulse amplitude modulated (PAM), where $u_n \in \left\{0,\frac{1}{M-1},\frac{2}{M-1},\dots,1\right\}$ for $n=0,1,\dots,N-1$. The complexity of HCM is the same as OFDM since an $N$-size FWHT also has a computational complexity of order $N \log_2 N$.

    \begin{figure} [!t]
    \begin{center}
    {\includegraphics[width=3.0in]{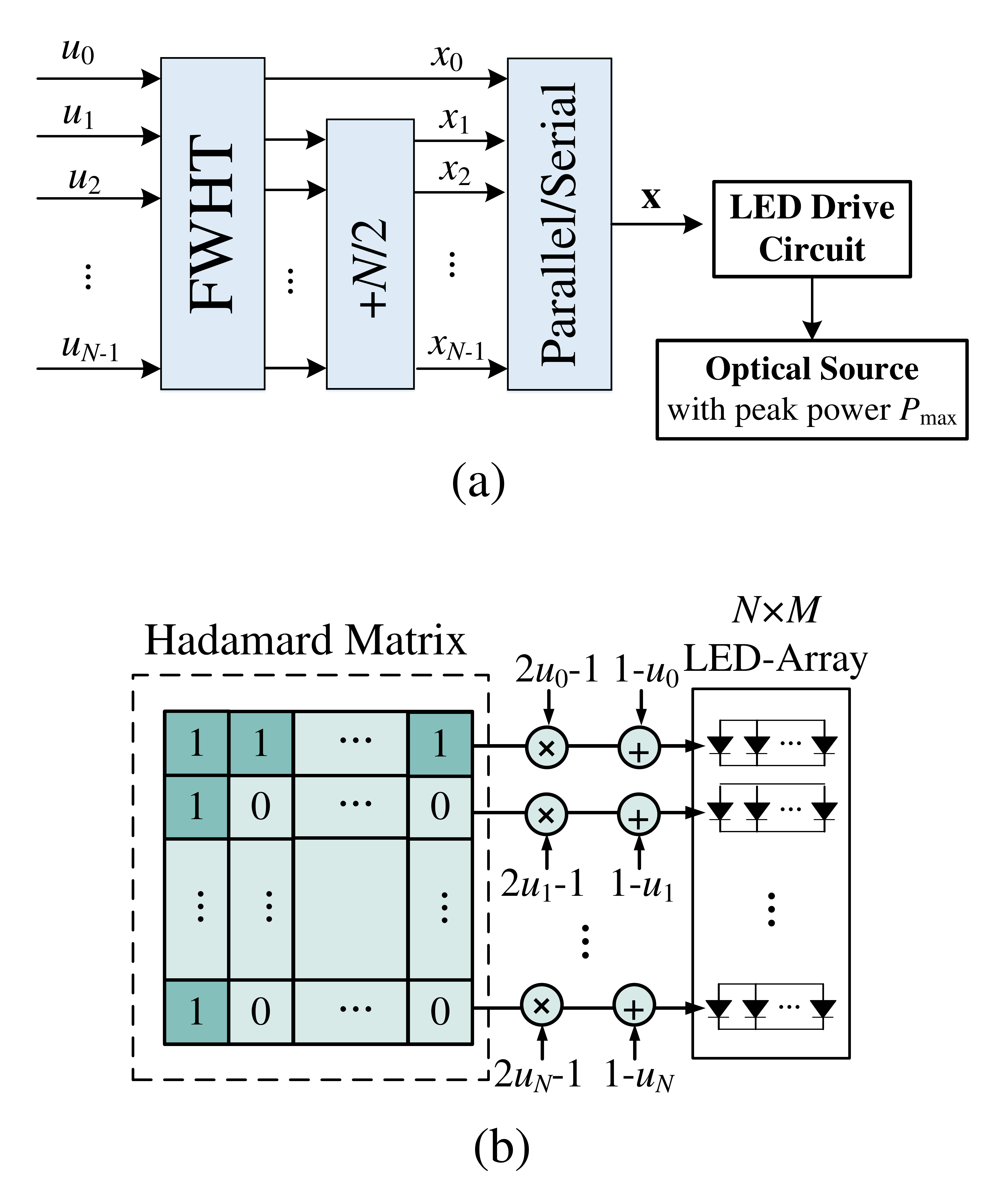}}
    \end{center}
    \vspace{-0.2in}
    \caption{Block diagram of the HCM transmitter using FWHT.}
    \label{Fig:HCM-Transmitter}
    \vspace{-0.2in}
    \end{figure}
    \begin{figure} [!b]
    \vspace{-0.25in}
    \begin{center}
    {\includegraphics[width=2.8in]{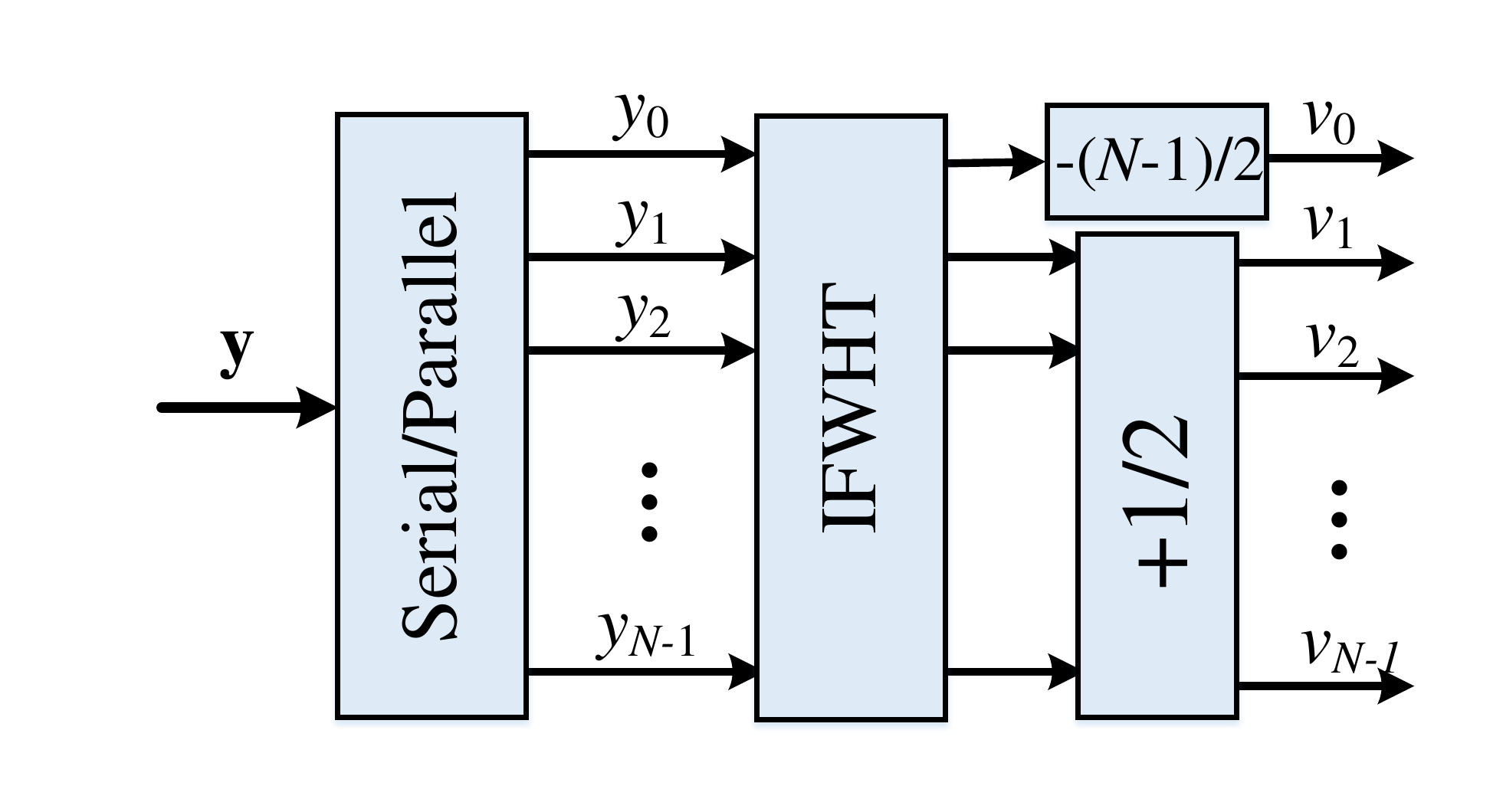}}
    \end{center}
    \vspace{-0.2in}
    \caption{Block diagram of the HCM receiver using IFWHT.}
    \label{Fig:HCM-Receiver}
    \end{figure}
%

Similar to \cite{Multilevel-EPPM12}, two structures can be used for the HCM transmitter. In the first structure, shown in Fig.~\ref{Fig:HCM-Transmitter}-(a), the HCM symbols generated are sent to an amplitude modulator that then modulates the optical source. This structure, which we call the single-source structure, can be used with the power-line communication (PLC) integrated VLC networks, where the data is send to the LED bulbs via the power lines and each component of the LED array cannot be modulated separately. In the single-source structure, as mentioned earlier in Section~\ref{sec:LED Nonlinearity}, the nonlinear transfer function of the optical source causes unequal spacing between the transmitted power levels, which makes the symbols more susceptible to noise, and therefore, a predistorter is required to make the power levels equal. A control circuit is also needed to compensate for the drift due to the thermal changes, which leads to an increased complexity of the transmitter.

In the second structure, the nonlinearity problem of the optical sources is solved by using each LED in an array in its on or off mode. The second structure, which is referred to as the LED-array structure, directly modulates a set of LEDs with one Hadamard code as shown in Fig.~\ref{Fig:HCM-Transmitter}-(b). Given that the $u_n$'s are $M$-PAM modulated, the LED-array structure uses a total of $N\times (M-1)$ LEDs to modulated the data, i.e., $(M-1)$ LEDs for each Hadamard code. This structure can be used in VLC systems in which each LED can be modulated independently. The LED-array structure guarantees equal spacing between the output optical power levels, and avoids the effect of the LED nonlinearity on the transmitted optical signal. Using HCM this structure is able to transmit only average powers less than $P_{\max}/2$, where $P_{\max}$ is the peak optical power of the LED array.

In either case, the decoded vector $\mathbf{v}$ is obtained from the received vector $\mathbf{y}$ as
    \begin{align}\label{HCM Decoder}
        \mathbf{v} = \frac{1}{N} \Big( \mathbf{H}{^\textmd{T}_N} \mathbf{y} - \mathbf{\overline{H}}{^\textmd{T}_N} \mathbf{y} \Big) + \frac{P}{2}[1-N,1,1\dots,1]^{\textmd{T}} ,
    \end{align}
which can be realized by an inverse FWHT (IFWHT) as shown in Fig.~\ref{Fig:HCM-Receiver}.
The noise due to the channel, $\mathbf{n}$, is assumed to be an additive white Gaussian noise (AWGN) vector with auto-covariance matrix $\sigma{^2_N}\mathbf{I}$, and hence, in the absence of nonlinearity the output signal of a non-dispersive channel, i.e., $h_{\ell}=0$ for $\ell \neq 0$, is given by $\mathbf{y}=\left( P/N \right) \mathbf{x}+\mathbf{n}$, where $P$ is the unclipped peak transmitted power.
Assuming a photo-detector with responsivity equal to 1, the decoded data can be rewritten as
    \begin{align}\label{Decoded Signal with Equivalent Noise}
        \mathbf{v} = \frac{P}{N}  \mathbf{u} + \mathbf{\tilde{n}},
    \end{align}
where $\mathbf{\tilde{n}} =  \frac{1}{N} \Big( \mathbf{H}{^\textmd{T}_N} - \mathbf{\overline{H}}{^\textmd{T}_N}\Big) \mathbf{n} $ is a $N \times 1$ noise vector with independent components.



%
    \begin{figure} [!b]
   \vspace{-0.2in}
    \begin{center}
    {\includegraphics[width=2.6in]{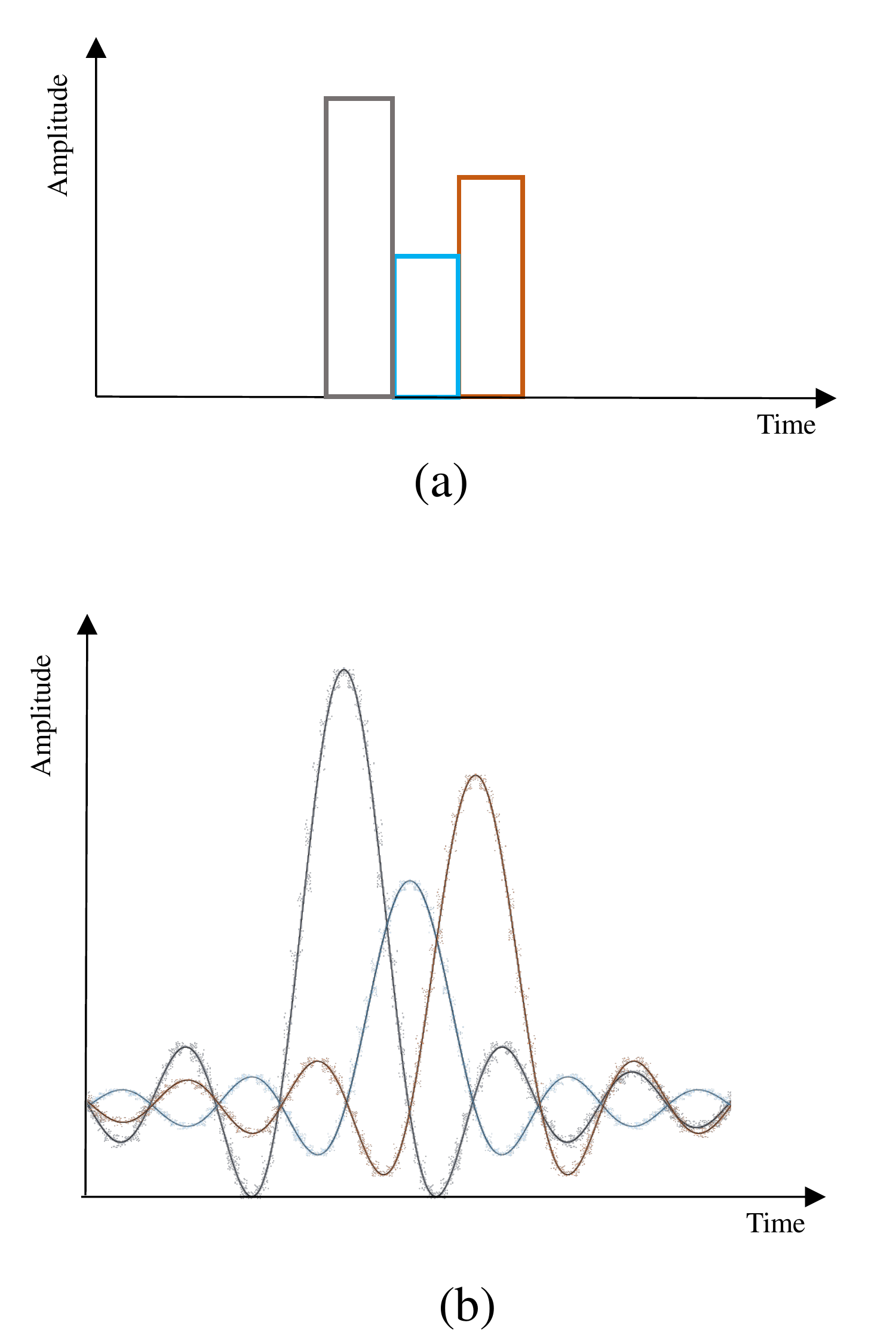}}
    \end{center}
    \vspace{-0.15 in}
    \caption{(a) An HCM signal, and (b) the transmitted pulses using sinc pulse-shaping.}
    \label{Fig:HCM-Sinc}
    \end{figure}
\subsection{Pulse Shaping to Increase Spectral Efficiency}
In practice, transmitting rectangular pulses requires a large bandwidth and is not spectrally efficient. In order to overcome this problem, we use sinc pulses instead of rectangular ones to transmit data. But since negative signals cannot be sent over the optical link, we add a DC bias to the signals to make them positive. Fig.~\ref{Fig:HCM-Sinc}-(b) illustrates the transmitted pulses for the three rectangular pulses shown in Fig.~\ref{Fig:HCM-Sinc}-(a). Replacing rectangular pulses with sinc pulses reduces the SNR by 0.83 dB.

\subsection{BER Calculation}

Through (\ref{Decoded Signal with Equivalent Noise}), the BER of $M$-PAM HCM can be calculated from \cite{BER-QAM-02} as
    \begin{align}\label{BER Expression for HCM}
        \textmd{BER}_{\textmd{HCM}} \approx \frac{(M-1)}{M\log_2 M} Q\Bigg( \sqrt{\frac{3}{\gamma(M^2-1)} \frac{P{^2}/N}{\sigma{^2_N}+\sigma{^2_{\text{clip}}}}  } \Bigg) .
    \end{align}
where $\gamma$ represents the penalty in SNR due to the pulse-shaping, which is 1.21 in this work, $\sigma{^2_N}$ is the variance of the additive Gaussian noise at the receiver and $\sigma{^2_{\text{clip}}}$ is the variance of the clipping noise.

For the LED-array transmitter structure in Fig.~\ref{Fig:HCM-Transmitter}-(b), $\sigma{^2_{\text{clip}}}=0$ and no further analysis is needed.
Since each of the $N-1$ columns of $\mathbf{H}{_N}$ that are used to modulate the data has an equal number of zeros and ones, HCM signals have a PAPR of 2, and therefore, for the single-source transmitter structure in Fig.~\ref{Fig:HCM-Transmitter}-(a), its signals are not clipped by LEDs for average power levels less than $P_{\max}/2$ and the clipping noise is zero, i.e., $\sigma{^2_{\text{clip}}}=0$.
For average powers larger than $P_{\max}/2$, we use (\ref{Variance of Clipping Distortion}) to find $\sigma{^2_{\text{clip}}}$. In order to find the pdf of $\mathbf{x}$, we first consider $\mathbf{u}$ to be a binary vector, and then we generalize the results to the case when the components of $\mathbf{u}$ are $M$-PAM.

For the binary case, $x_n \in \{0,1,2,\dots,N-1,N\}$ for $n=1,2,\dots,N$, and the probability that $x_n=k$ is equal to
    \begin{align}\label{}
        {\text{Pr}}\left(x_n=k\right) = {\binom{N}{k}} \left(\frac{1}{2}\right)^N, \hspace{0.3cm} k=0,1,\dots,N.
    \end{align}
Through (\ref{Variance of Clipping Distortion}), the clipping noise for binary HCM is
    \begin{align}\label{}
        \sigma{^2_{\text{clip}}} = \left(\frac{1}{2}\right)^N \sum_{k=\lceil N\frac{P_{\max}}{P}\rceil}^N \left(k\frac{P}{N}-P_{\max}\right)^2 {\binom{N}{k}} .
    \end{align}
where $\lceil x \rceil$ is the smallest integer larger than $x$.

For $M$-PAM HCM, the components of $\mathbf{x}$ take values from a larger set as $x_n \in \{0,\frac{1}{M-1},\frac{2}{M-1},\dots,N\}$ for $n=1,2,\dots,N$, and the probability of $x_n=\frac{k}{M-1}$ is
    \begin{align}\label{}
        {\text{Pr}}\left(x_n=\frac{k}{m-1}\right) = C(m,N,k) \left(\frac{1}{m}\right)^N, \hspace{0.3cm} k=0,1,\dots,N.
    \end{align}
where $C(m,N,k)$'s are the \emph{extended binomial coefficients} defined as the coefficients of $x^k$ in the expansion of $(1+x+x^2+\dots,x^{(m-1)})^N$ \cite{Extended-Binom-13}.

    \begin{figure} [!b]
   \vspace{-0.2 in}
    \begin{center}
    {\includegraphics[width=3.4in]{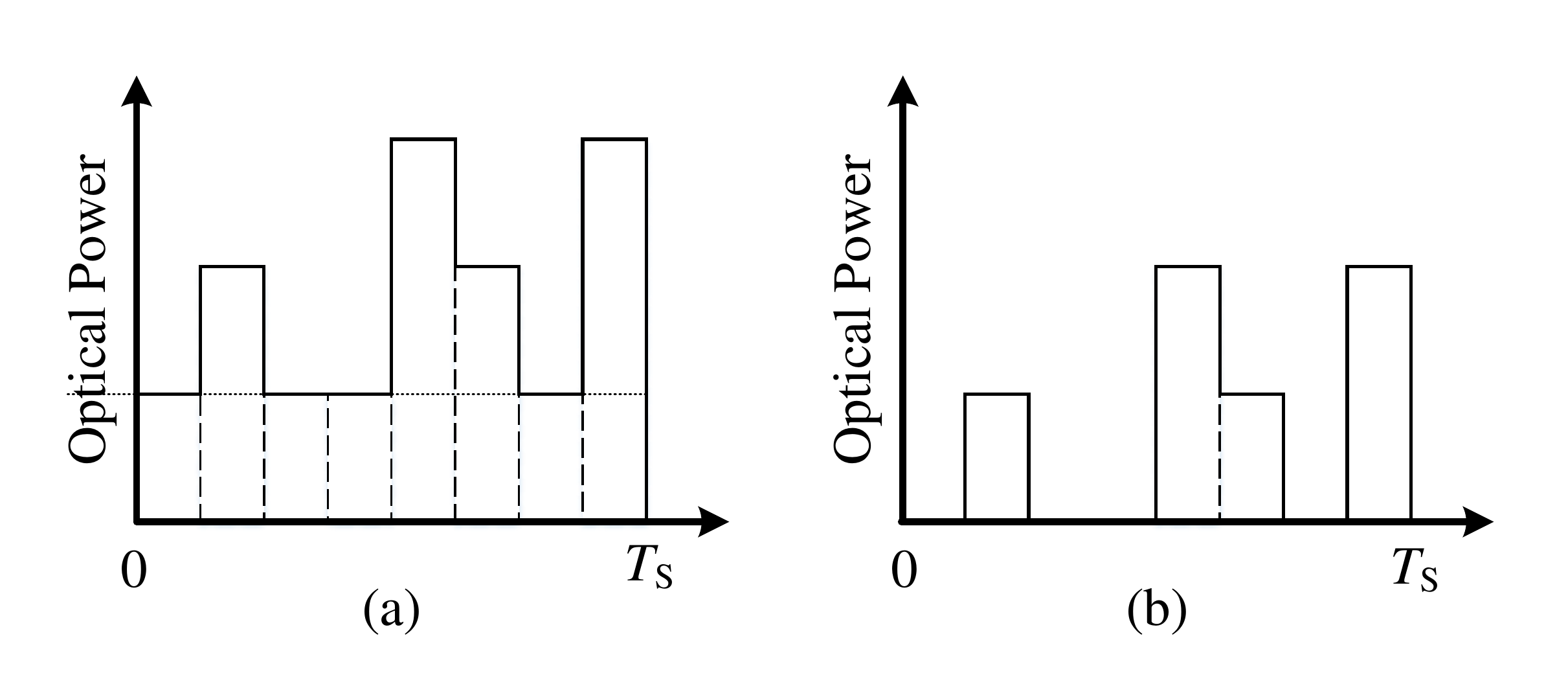}}
    \end{center}
    \vspace{-0.25 in}
    \caption{(a) An HCM signal, and (b) its corresponding DC reduced signal.}
    \label{Fig:DC-Removal}
    \end{figure}
    \begin{figure} [!t]
     \vspace{-1.8in}
    \hspace{-0.5in}{\includegraphics[width=4.6in]{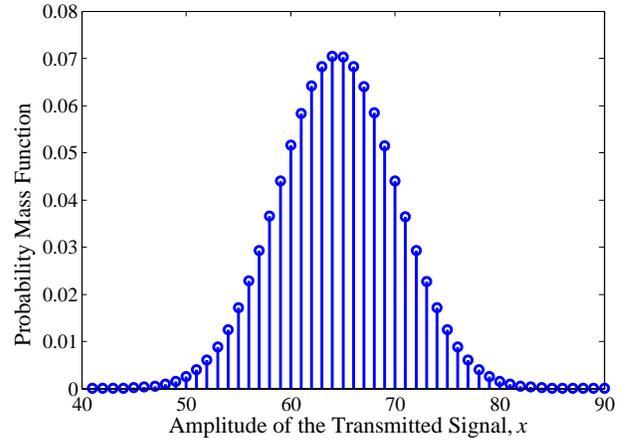}}
    \vspace{-2.0in}
    \begin{center}
    {\small (a)}
    \end{center}
    \vspace{-1.7in}
    \hspace{-0.5in}\includegraphics[width=4.6in]{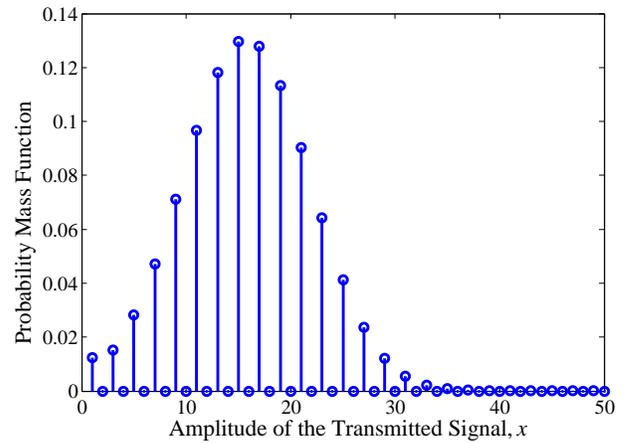}
    \vspace{-2.0in}
    \begin{center}
    {\small (b)}
    \end{center}
    \vspace{-0.15 in}
    \caption{Probability mass function for (a) binary HCM, and (b) binary DCR-HCM.}
    \label{Fig:pdf}
    \vspace{-0.25 in}
    \end{figure}

\subsection{Increasing Energy Efficiency Using DC-Reduced HCM} \label{sec:Removing DC Bias}
As shown in \cite{HCM-Globecom-14}, the DC part of HCM signals (before the pulse-shaping) can be reduced without losing any information, making HCM more average power efficient. This is important when for illumination reasons the light should be dimmed, i.e., not operated at its brightest level. Let the first component of $\mathbf{u}$ be set to zero and only $N-1$ codewords of the Hadamard matrix be modulated, as proposed in \cite{HCM-Globecom-14}. In this scheme, which is called DC-reduced HCM (DCR-HCM), the average transmitted power is reduced by sending $(\mathbf{x} - \min \mathbf{x}) $ instead of $\mathbf{x}$. The reduced DC level is per HCM symbol and its value can be different for each symbol. The same receiver structure as in Fig.~\ref{Fig:HCM-Receiver} can be used to decode the received signals.
Fig.~\ref{Fig:DC-Removal} shows an example of DC reduction in an HCM symbol of size $N=8$, where the transmitted energy of the HCM symbol in Fig.~\ref{Fig:DC-Removal}-(a) is reduced by a factor of $3/7$ in its corresponding DCR-HCM symbol in Fig.~\ref{Fig:DC-Removal}-(b). This technique can only be easily implemented for the single-source transmitter structure, and an intermediate circuit is required to apply this technique to the LED-array structure.

The DC reduction technique decreases the probability of large amplitudes of $\mathbf{x}$, which makes the signals less likely to be clipped by the optical source, and therefore, DCR-HCM can achieve lower BERs at lower average power levels compared to HCM in peak-power limited systems. The probability mass function of the transmitted signal, ${\text{Pr}}\left(x=k\right)$, is plotted using Monte-Carlo simulation for binary HCM and DCR-HCM respectively in Fig.~\ref{Fig:pdf}-(a) and Fig.~\ref{Fig:pdf}-(b) for $N=128$. According to these results, the peak of ${\text{Pr}}\left(x=k\right)$ is shifted to lower $x$'s for DCR-HCM and the high amplitudes in DCR-HCM signals have lower probabilities compared to that of HCM \footnote{Note that the DC value could instead be increased to its maximum value for scenarios that require even brighter illumination. This idea is entirely analogous to DCR-HCM and is therefore not discussed further here.}.

\begin{figure} [!t]
   \vspace*{-1.5 in}
    \hspace{-0.4 in}{\includegraphics[width=4.4in]{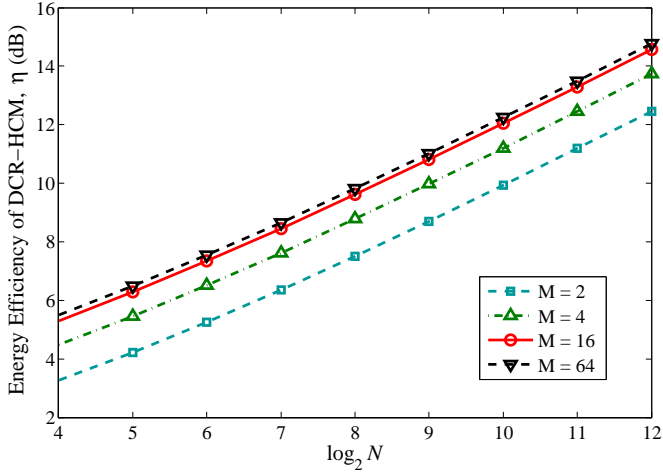}}
    \vspace*{-1.7 in}
    \caption{Energy efficiency of DCR-HCM versus the order of the Hadamard transform, $\log_2 N$, for $M=2$, 4, 16 and 64.}
    \label{Fig:DCR-HCM Energy Efficiency}
    \vspace{-0.15 in}
    \end{figure}

The energy efficiency of DCR-HCM, denoted $\eta$, is defined as
    \begin{align}\label{}
        \eta = \frac{ E\left\{ x_n \right\} }{ E\left\{ x_n \right\} - E \{\min \mathbf{x}\} },
    \end{align}
In this definition we have used the fact that all components of $\mathbf{x}$ have the same mean, i.e., $E\{x_n\}$ is fixed for all $n=0,1,\dots,N-1$.
Fig.~\ref{Fig:DCR-HCM Energy Efficiency} shows $\eta$ as a function of the order of the Hadamard transform. In this figure, the data sequence is assumed to be $M$-ary PAM modulated. According to these results, the energy efficiency of DCR-HCM increases almost linearly with $\log_2 N$, and also increases slightly with increasing $M$.

\subsection{Dispersive Channels} \label{sec:Interleaved HCM}
VLC experience dispersive channels that create inter-symbol interference (ISI) on the transmitted signals, and therefore, any practical modulation technique must be resistant against ISI. In OFDM, a cyclic prefix is used as a guard interval in order to eliminate the intersymbol interference from adjacent symbols. It also allows the linear convolution of a frequency-selective multipath channel to be modelled as a circular convolution by repeating the end of the symbol, which simplifies channel estimation and equalization at the receiver. Likewise, a cyclic prefix is used for HCM symbols to avoid interference from other symbols, and therefore, the interference on a symbol is intra-symbol interference that is caused by its own pulses. This also allows us to use cyclic shifts of transmitted vectors instead of their right shifts in our analysis. Under these assumptions, given $\mathbf{x}$ is sent, and ignoring background light and other noise, the received signal is proportional to the nonlinearly distorted (clipped) version of $\sum\limits_{\ell} {h_{\ell}} \mathbf{x}{^{(\ell)}}$. In this section two techniques are proposed to handle the dispersive nature of the channel: interleaving and equalization.

\begin{figure} [!t]
    \hspace{-0.1in}{\includegraphics[width=3.7in]{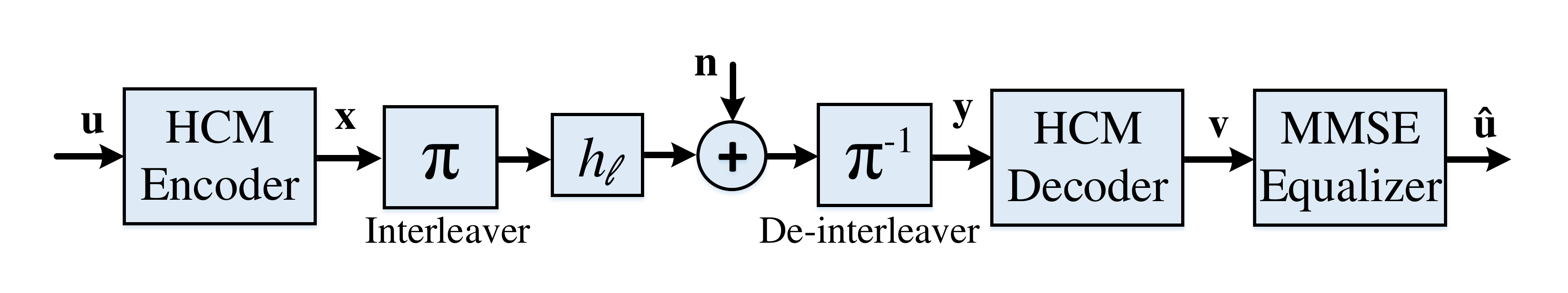}}
    \vspace{-0.25in}
    \caption{Schematic view of an interleaved HCM system using an for dispersive channels.}
    \label{Fig:Interleaved-HCM}
    \vspace{-0.25in}
    \end{figure}

Hadamard matrices consist of rows that are cyclic shifts, which increases the similarity between Hadamard codewords in dispersive channels and makes HCM vulnerable to ISI. Interleaving is an effective solution that reduces the ISI by decreasing the cross-correlation of the codewords with their cyclic shifts \cite{VLC-JLT-I-13}.
In this technique, as shown in Fig.~\ref{Fig:Interleaved-HCM}, a symbol-length interleaver and a deinterleaver are used at the transmitter and receiver, respectively, to reduce the effects of intra-symbol ISI due to a dispersive channel. The interleaver is a permutation matrix, $\mathbf{\pi}$, and the deinterleaver is its inverse, $\mathbf{\pi}^{-1}$. Hence, $\mathbf{x} \mathbf{\pi}$ is sent instead of $\mathbf{x}$. For a non-equalizing receiver, the best interleaver matrix is the one that evenly distributes the interference over all symbols, and can be found using binary linear programming \cite{VLC-JLT-I-13}. In non-dispersive channels, the performance of interleaved HCM is the same as HCM since $\mathbf{\pi} \mathbf{\pi}^{-1}=\mathbf{I}$.

%

Assuming $\mathbf{x}$ is transmitted, the noiseless output of the channel is proportional to $\sum\limits_{\ell} {h_{\ell}} \left( \pi \mathbf{x}\right)^{(\ell)}$. Then the decoded signal can be written as
    \begin{align}\label{Received Signal with Interleaver}
        \mathbf{v} = &\frac{ P}{N} \sum_{\ell} {h_{\ell}} \Big(\mathbf{H}{_N} - \overline{\mathbf{H}}{_N} \Big) \pi^{-1} \mathbf{I}^{(\ell)} \pi \Big( \mathbf{u}  \mathbf{H}{_N} + \overline{\mathbf{u}} \overline{\mathbf{H}}{_N} \Big) \nonumber\\ &+ \frac{P}{2}[1-N,1,1\dots,1]^{\textmd{T}} + \mathbf{\tilde{n}}.
    \end{align}
%
Defining $\mathbf{G}:= \sum\limits_{\ell} {h_{\ell}} \mathbf{I}^{(\ell)} $,
(\ref{Received Signal with Interleaver}) can be written as
    \begin{align}\label{Interference on the Received Signal with Interleaver - Simplified}
        \mathbf{v} = P \Big( \mathbf{H}{_N} - \overline{\mathbf{H}}{_N} \Big) \pi^{\text{T}} \mathbf{G} \pi \Big(\mathbf{H}{_N} - \overline{\mathbf{H}}{_N} \Big)\frac{\left( 2\mathbf{u} -1 \right)}{2N} + \frac{P}{2} + \mathbf{\tilde{n}}.
    \end{align}

In addition, minimum mean square error (MMSE) equalization is an effective technique to estimate the data at the output of a noisy dispersive channel. The goal in this equalizer is to find the matrix $\mathbf{W}$ that minimizes the trace of $E\left\{ (\mathbf{u}-\hat{\mathbf{u}}) (\mathbf{u}-\hat{\mathbf{u}})^{\textmd{T}} \right\}$, where $\hat{\mathbf{u}}= \mathbf{W} (\mathbf{v}-P/2) + 1/2$ is an estimate of the data sent, $\mathbf{u}$, based on the observation of the decoded vector $\mathbf{v}$. According to \cite{Proakis}, the optimum $\mathbf{W}$ is given by
    \begin{align}\label{Optimal W Matrix for MMASE}
        \mathbf{W}= \mathbf{C}_{\mathbf{u}\mathbf{v}} \mathbf{C}{_{\mathbf{v}}^{-1}},
    \end{align}
where $\mathbf{C}_{\mathbf{u}\mathbf{v}}$ is cross-covariance matrix between $\mathbf{u}$ and $\mathbf{v}$, and $\mathbf{C}{_{\mathbf{v}}}$ is auto-covariance matrix of $\mathbf{v}$. For (\ref{Optimal W Matrix for MMASE}), the error is
    \begin{align}\label{MMSE Error}
        \text{LMMSE}=\text{tr}\left\{ \mathbf{C}_{\mathbf{u}} - \mathbf{C}_{\mathbf{u}\mathbf{v}} \mathbf{C}{_{\mathbf{v}}^{-1}} \mathbf{C}_{\mathbf{v}\mathbf{u}} \right\}.
    \end{align}

For HCM, using (\ref{HCM Encoder}) and (\ref{HCM Decoder}) we get
    \begin{align}\label{}
        \mathbf{C}_{\mathbf{u}\mathbf{v}} = \frac{P}{4N} \Big( \mathbf{H}{_N} - \overline{\mathbf{H}}{_N} \Big) \pi^{\text{T}} \mathbf{G} \pi \Big(\mathbf{H}{_N} - \overline{\mathbf{H}}{_N} \Big),
    \end{align}
and
    \begin{align}\label{}
        \mathbf{C}_{\mathbf{v}} = \frac{\sigma{_N^2}}{N} \mathbf{I} + \frac{P{^2}}{4 N}  \Big( \mathbf{H}{_N} - \overline{\mathbf{H}}{_N} \Big) \pi^{\text{T}} \mathbf{G} \mathbf{G}^{\textmd{T}} \pi \Big(\mathbf{H}{_N} - \overline{\mathbf{H}}{_N} \Big) ,
    \end{align}
where we have used the symmetry property of Hadamard matrices, i.e., $\mathbf{H}_N = \mathbf{H}{^\textmd{T}_N}$.


%
    \begin{figure} [!t]
   \vspace*{-1.2 in}
    \hspace{-0.2 in}{\includegraphics[width=3.7in]{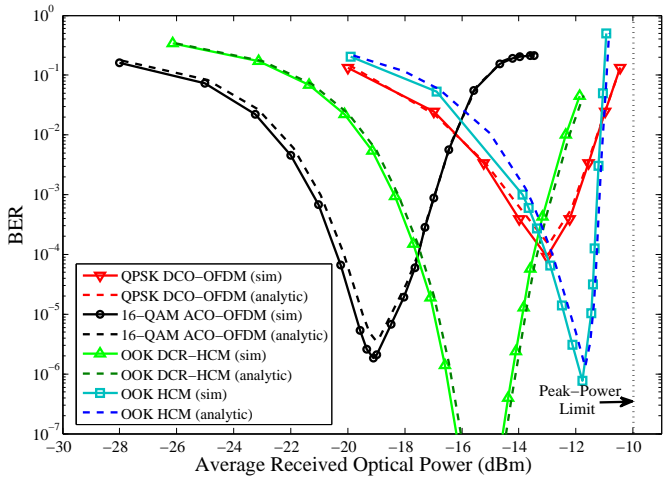}}
    \vspace*{-1.3 in}
    \caption{Analytical (dashed lines) and simulated (solid lines) BER of HCM, DRC-HCM, ACO-OFDM and DCO-OFDM versus the average received power in an ideal channel, i.e., $h_0=1$.}
    \label{Fig:BER-of-OFDM-vs-HCM}
    \vspace{-0.2 in}
    \end{figure}
    \begin{figure} [!b]
   \vspace*{-1.45 in}
    \hspace{-0.1 in}{\includegraphics[width=3.8in]{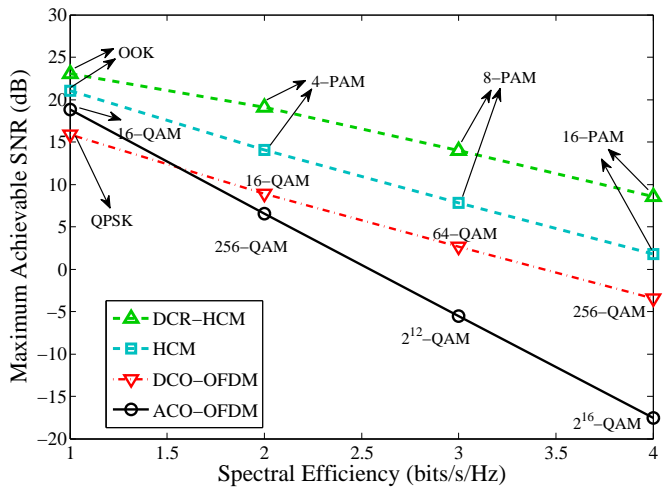}}
    \vspace*{-1.4 in}
    \caption{Maximum achievable SNR versus the spectral efficiency for HCM, DCR-HCM, ACO-OFDM and DCO-OFDM.}
    \label{Fig:SNR-vs-Spectral-Efficiency}
    \end{figure}
    \begin{figure} [!t]
   \vspace*{-1.4 in}
    \hspace{-0.2 in}{\includegraphics[width=4.0in]{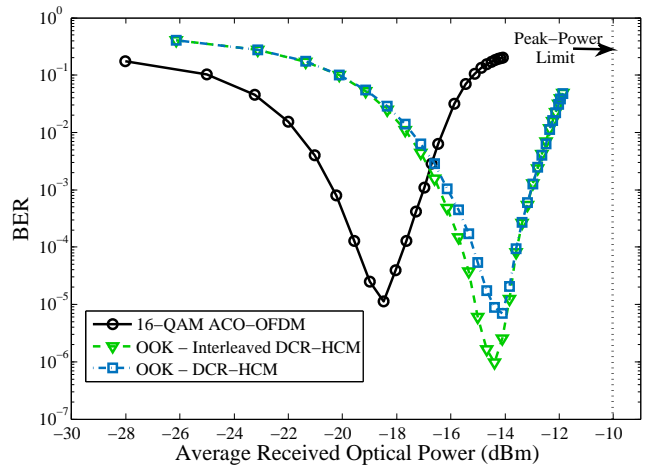}}
    \vspace*{-1.55 in}
    \caption{Simulated BER of ACO-OFDM, DCR-HCM, and interleaved DCR-HCM in a dispersive channel with impulse response $h_0=0.5$, $h_1=0.3$ and $h_2=0.2$ using an MMSE equalizer for DCR-HCM and one-tap equalizer for ACO-OFDM.}
    \label{Fig:MMSE}
    \vspace{-0.25 in}
    \end{figure}

\section{Numerical Results} \label{sec:Numerical Results}
In this section, numerical results using simulation and  analysis are presented to compare the performance of HCM and DCR-HCM to ACO-OFDM and DCO-OFDM.
In the simulations, the sources are assumed to be ideal peak-power limited sources as shown in Fig.~\ref{Fig:LED-Transfer Function}-(b) and the peak received power is assumed to be 0.1 mW. The optical source is modulated as in Fig.~\ref{Fig:HCM-Transmitter}-(a). The transmitter and receiver are assumed to be perfectly synchronized. These results are generated using sinc pulse-shaping, shown in Fig.~\ref{Fig:HCM-Sinc}-(b).

The BERs of HCM, DCR-HCM, ACO-OFDM and DCO-OFDM are plotted versus the average received optical power in Fig.~\ref{Fig:BER-of-OFDM-vs-HCM} for $N=128$. The parameters of these techniques are chosen such that all have spectral efficiency of 1, i.e., the same data-rate.  The analytical BERs are in good agreement with Monte-Carlo simulation results. For HCM and DCR-HCM, (\ref{BER Expression for HCM}) and (\ref{Variance of Clipping Distortion}) are used to plot the analytical results, while those of ACO-OFDM and DCO-OFDM are plotted using the results of \cite{Optical-OFDM-Clipping-Hass-11, Optical-OFDM-Armstrong-13}.
ACO-OFDM and DCO-OFDM use 16-QAM and QPSK, respectively, to modulate the data, while on-off keying (OOK) is used to modulate the data in HCM and DCR-HCM. The results are plotted assuming an AWGN channel for a noise level of $\sigma{^2_N}=2$ $\mu$W. The BER of ACO-OFDM decreases by increasing the average optical power until it reaches an optimum point, and increases afterwards due to the clipping imposed distortion. DCO-OFDM has the same behavior but for higher average powers. The average optical power of DCO-OFDM is proportional to its DC level, and hence, it reaches its lowest BER for an average power of half of the peak power. According to these results, DCO-OFDM, HCM and DCR-HCM are better choices for VLC systems since they have a better performance at high average optical powers. As one can see, HCM and DCR-HCM can achieve lower BERs compared to OFDM techniques. Between these two, DCR-HCM uses signals with lower amplitudes and is therefore more resistant to clipping noise.

The maximum {\em achievable SNR } is plotted versus the spectral efficiency for HCM, DCR-HCM, ACO-OFDM and DCO-OFDM for a noise level of $\sigma{^2_N}=0.5$ $\mu$W. Analytical BER expressions of the form $\alpha Q\left( \sqrt{\text{SNR}} \right)$ taken from (\ref{BER Expression for HCM}) for HCM and DCR-HCM and from \cite{Optical-OFDM-Clipping-Hass-11, Optical-OFDM-Armstrong-13} for ACO-OFDM and DCO-OFDM are used to plot these results. For each technique, the optimum average power that minimizes the BER is found analytically and then the SNR is calculated for that average power. For DCO-OFDM, the DC level is set to $P_{\max}/2$ since it minimizes the clipping noise, and hence, corresponds to the maximum SNR among all possible DC levels. Monte-Carlo simulation is used to find the probability mass function of DCR-HCM, and then it is used in (\ref{Variance of Clipping Distortion}) to find the clipping noise power. Note that ACO-OFDM is only better than DCO-OFDM at low spectral efficiencies, and it requires high-order QAM to get spectral efficiencies larger than 2, which is impractical. Hence, ACO-OFDM is not suitable for spectrally efficient VLC systems. HCM and DCR-HCM are able to provide a higher achievable SNR for all spectral efficiencies tested.

The simulated BER of 16-QAM ACO-OFDM is compared to that of OOK DCR-HCM in Fig.~\ref{Fig:MMSE} for a highly dispersive channel with a discrete-time equivalent impulse response of ${h_0}=0.4$, $h_1=0.3$ and ${h_2}=0.3$ (using $N=128$ samples per symbol) using an MMSE equalizer for DCR-HCM and one-tap equalizer for ACO-OFDM. Both techniques use a cyclic prefix of length 4. The noise is assumed to be $\sigma{^2_N}=6$ $\mu$W. According to these results, DCR-HCM can achieve lower BERs compared to ACO-OFDM. The BER of the system using MMSE equalization on the interleaved DCR-HCM is also plotted versus the average optical power. The interleaver is found using the binary linear program described in \cite{VLC-JLT-I-13}. As shown in Fig.~\ref{Fig:MMSE}, interleaving improves the performance of the MMSE equalizer by almost an order of magnitude by spreading the interference equivalently on all Hadamard codewords.


\section{Conclusion} \label{sec:Conclusion}
In this paper, HCM and its modified form DCR-HCM are proposed as alternative techniques to OFDM for LED-based VLC systems that require high illumination levels. HCM and DCR-HCM achieve lower BERs compared to ACO-OFDM for high average power since they transmit signals with lower peak amplitudes.  The energy efficiency of HCM can be improved by reducing the DC part of the transmitted signals without losing any information. This efficiency is shown to increase with the size of the Hadamard transform and the size of the PAM constellation used.
The performance of HCM and DCR-HCM is shown to surpass that of ACO-OFDM and DCO-OFDM, and they are able to achieve 2 to 3 orders of magnitude lower BERs.
Interleaving along with MMSE equalization can effectively decrease the BER of HCM by an order of magnitude in dispersive VLC channels.

\bibliographystyle{IEEEtran}
\balance
\bibliography{EPPM,Thesis,VLC,OFDM,Coded-Multiplexing}

\begin{thebibliography}{10}
\providecommand{\url}[1]{#1}
\def\UrlFont{\rmfamily}
\providecommand{\newblock}{\relax}
\providecommand{\bibinfo}[2]{#2}
\providecommand\BIBentrySTDinterwordspacing{\spaceskip=0pt\relax}
\providecommand\BIBentryALTinterwordstretchfactor{4}
\providecommand\BIBentryALTinterwordspacing{\spaceskip=\fontdimen2\font plus
\BIBentryALTinterwordstretchfactor\fontdimen3\font minus
  \fontdimen4\font\relax}
\providecommand\BIBforeignlanguage[2]{{%
\expandafter\ifx\csname l@#1\endcsname\relax
\typeout{** WARNING: IEEEtran.bst: No hyphenation pattern has been}%
\typeout{** loaded for the language `#1'. Using the pattern for}%
\typeout{** the default language instead.}%
\else
\language=\csname l@#1\endcsname
\fi
#2}}

\bibitem{VLC-Ghassemlooy}
Z.~Ghassemlooy, W.~Popoola, and S.~Rajbhandari, \emph{Optical Wireless
  Communications: System and Channel Modelling with {MATLAB}}.\hskip 1em plus
  0.5em minus 0.4em\relax {CRC} Press, 2012.

\bibitem{OFDM-Book}
R.~Prasad, \emph{{OFDM} for Wireless Communications Systems}.\hskip 1em plus
  0.5em minus 0.4em\relax Artech House, Inc., 2004.

\bibitem{Optical-OFDM-Armstrong-13}
S.~Dissanayake and J.~Armstrong, ``Comparison of {ACO-OFDM}, {DCO-OFDM} and
  {ADO-OFDM} in {IM/DD} systems,'' \emph{J. Lightw. Technol.}, vol.~31, no.~7,
  pp. 1063--1072, April 2013.

\bibitem{ACO-OFDM-06}
J.~Armstrong and A.~Lowery, ``Power efficient optical {OFDM},'' \emph{Electron.
  Lett.}, vol.~42, no.~6, pp. 370--372, March 2006.

\bibitem{DCO-OFDM-06}
O.~Gonzalez, R.~Perez-Jimenez, S.~Rodriguez, J.~Rabadan, , and A.~Ayala,
  ``Adaptive {OFDM} system for communications over the indoor wireless optical
  channel,'' \emph{IEE Proceedings-Optoelectronics}, vol. 153, pp. 139--144,
  2006.

\bibitem{ADO-OFDM-11}
S.~Dissanayake, K.~Panta, and J.~Armstrong, ``A novel technique to
  simultaneously transmit {ACO-OFDM} and {DCO-OFDM} in {IM/DD} systems,''
  \emph{Proc. of IEEE Global Telecommun. Conf., ({GLOBECOM})}, pp. 782--786,
  Dec 2011.

\bibitem{OFDM-PAPR-Reduction-VLC-Zabih-14}
W.~Popoola, Z.~Ghassemlooy, and B.~Stewart, ``Pilot-assisted {PAPR} reduction
  technique for optical {OFDM} communication systems,'' \emph{J. Lightw.
  Tech.}, vol.~32, no.~7, pp. 1374--1382, April 2014.

\bibitem{OFDM-Hadamard-11}
M.~Ahmed, S.~Boussakta, B.~Sharif, and C.~Tsimenidis, ``{OFDM} based on low
  complexity transform to increase multipath resilience and reduce {PAPR},''
  \emph{IEEE Trans. Signal Process.}, vol.~59, no.~12, pp. 5994--6007, 2011.

\bibitem{OFDM-Hadamard-12}
J.~Xiao, J.~Yu, X.~Li, Q.~Tang, H.~Chen, F.~Li, Z.~Cao, and L.~Chen,
  ``{Hadamard} transform combined with companding transform technique for
  {PAPR} reduction in an optical direct-detection {OFDM} system,'' \emph{IEEE
  J. Opt. Commun. Netw.}, vol.~4, no.~10, pp. 709--714, Oct 2012.

\bibitem{OFDM-Hadamard-BER-03}
Y.-P. Lin and S.-M. Phoong, ``{BER} minimized {OFDM} systems with channel
  independent precoders,'' \emph{IEEE Trans. Signal Process.}, vol.~51, no.~9,
  pp. 2369--2380, 2003.

\bibitem{OFDM-Hadamard-Fading-10}
S.~Wang, S.~Zhu, and G.~Zhang, ``A {Walsh-Hadamard} coded spectral efficient
  full frequency diversity {OFDM} system,'' \emph{IEEE Trans. Commun.},
  vol.~58, no.~1, pp. 28--34, January 2010.

\bibitem{OFDM-for-VLC-Wang-14}
Z.~Wang, Q.~Wang, S.~Chen, and L.~Hanzo, ``An adaptive scaling and biasing
  scheme for {OFDM}-based visible light communication systems,'' \emph{Opt.
  Exp.}, vol.~22, no.~10, pp. 12\,707--12\,715, May 2014.

\bibitem{Dimming-for-OFDM-13}
H.~Elgala and T.~D.~C. Little, ``Reverse polarity optical-{OFDM} {(RPO-OFDM)}:
  dimming compatible {OFDM} for gigabit {VLC} links,'' \emph{Opt. Express},
  vol.~21, no.~20, pp. 24\,288--24\,299, 2013.

\bibitem{PWM-PPM-VLC-Zabih-14}
J.-h. Choi, E.-b. Cho, Z.~Ghassemlooy, S.~Kim, and C.~Lee, ``Visible light
  communications employing {PPM} and {PWM} formats for simultaneous data
  transmission and dimming,'' \emph{Optical and Quantum Electron.}, pp. 1--14,
  May 2014.

\bibitem{VLC-Networking13}
M.~Noshad and M.~Brandt-Pearce, ``High-speed visible light indoor networks
  based on optical orthogonal codes and combinatorial designs,'' \emph{Proc.
  IEEE Global Commun. Conf. ({GLOBECOM})}, Athlanta, GA, Dec. 2013.

\bibitem{EPPM12}
------, ``Expurgated {PPM} using symmetric balanced incomplete block designs,''
  \emph{IEEE Commun. Lett.}, vol.~16, no.~7, pp. 968--971, 2012.

\bibitem{Multilevel-EPPM12}
------, ``Multilevel pulse-position modulation based on balanced incomplete
  block designs,'' \emph{Proc. IEEE Global Commun. Conf. ({GLOBECOM})},
  Anaheim, CA, Dec. 2012.

\bibitem{HCM-Globecom-14}
------, ``Hadamard coded modulation: An alternative to {OFDM} for wireless
  optical communications,'' \emph{Proc. IEEE Global Commun. Conf.
  ({GLOBECOM})}, pp. 2102--2107, Austin, TX, Dec. 2014.

\bibitem{Hadamard-in-CDMA-04}
K.~Paterson, ``On codes with low peak-to-average power ratio for multicode
  {CDMA},'' \emph{IEEE Tran. Inf. Theory}, vol.~50, no.~3, pp. 550--559, March
  2004.

\bibitem{VLC-JLT-I-13}
M.~Noshad and M.~Brandt-Pearce, ``Application of expurgated {PPM} to indoor
  visible light communications - part {I}: Single-user systems,'' \emph{J.
  Lightw. Technol.}, vol.~32, no.~5, pp. 875--882, March 2014.

\bibitem{Illuminance-Standards}
D.~DiLaura, K.~Houser, R.G.Mistrick, and G.~Steffy, \emph{The Lighting
  Handbook, 10th Ed.}\hskip 1em plus 0.5em minus 0.4em\relax Illuminating
  Engineering Society of North America (IESNA), 2011.

\bibitem{VLC-Lighting-Requirements-13}
J.~Gancarz, H.~Elgala, and T.~Little, ``Impact of lighting requirements on
  {VLC} systems,'' \emph{IEEE Commun. Mag.,}, vol.~51, no.~12, pp. 34--41,
  December 2013.

\bibitem{LED-Efficiency-Limit-10}
Y.~Narukawa, M.~Ichikawa, D.~Sanga, M.~Sano, and T.~Mukai, ``White light
  emitting diodes with super-high luminous efficacy,'' \emph{Journal of Physics
  D: Applied Physics}, vol.~43, 2010.

\bibitem{Optical-OFDM-Nonlinear-LED-13}
S.~Dimitrov and H.~Haas, ``Information rate of {OFDM}-based optical wireless
  communication systems with nonlinear distortion,'' \emph{J. Lightw.
  Technol.}, vol.~31, no.~6, pp. 918--929, March 2013.

\bibitem{OWC-IR-Book-08}
R.~Ramirez-Iniguez, S.~M. Idrus, and Z.~Sun, \emph{Optical Wireless
  Communications: IR for Wireless Connectivity}.\hskip 1em plus 0.5em minus
  0.4em\relax CRC Press, 2008.

\bibitem{LED-Level-Quantization-Haas-13}
T.~Fath, C.~Heller, and H.~Haas, ``Optical wireless transmitter employing
  discrete power level stepping,'' \emph{J. Lightw. Tech.}, vol.~31, no.~11,
  pp. 1734--1743, June 2013.

\bibitem{Amp-Distortion-Bussgang}
J.~Bussgang, ``Cross correlation function of amplitude-distorted {Gaussian}
  signals,'' \emph{Research Laboratory for Electronics, Massachusetts Institute
  of Technology, Cambridge, MA, Technical Report 216}, Mar.

\bibitem{OWC-Channel05}
O.~González, S.~Rodriguez, R.~Perez-Jimenez, B.~Mendoza, and A.~Ayala, ``Error
  analysis of the simulated impulse response on indoor wireless optical
  channels using a {Monte Carlo}-based ray-tracing algorithm,'' \emph{IEEE
  Trans. Commun.}, vol.~53, no.~1, pp. 124--219, Jan. 2005.

\bibitem{OWC-Channel11}
K.~Lee, H.~Park, and J.~Barry, ``Indoor channel characteristics for visible
  light communications,'' \emph{IEEE Commun. Lett.}, vol.~15, no.~2, pp.
  217--219, Feb. 2011.

\bibitem{Hadamard-Book}
K.~J. Horadam, \emph{Hadamard Matrices and Their Applications}.\hskip 1em plus
  0.5em minus 0.4em\relax Princeton University Press, 2006.

\bibitem{BER-QAM-02}
K.~Cho and D.~Yoon, ``On the general {BER} expression of one- and
  two-dimensional amplitude modulations,'' \emph{IEEE Trans. Commun.}, vol.~50,
  no.~7, pp. 1074--1080, Jul 2002.

\bibitem{Extended-Binom-13}
S.~Eger, ``Restricted weighted integer compositions and extended binomial
  coefficients,'' \emph{Journal of Integer Sequences}, vol.~16, 2013.

\bibitem{Proakis}
J.~Proakis and M.~Salehi, \emph{Digital Communications, 5th Ed.}\hskip 1em plus
  0.5em minus 0.4em\relax McGraw-Hill, 2007.

\bibitem{Optical-OFDM-Clipping-Hass-11}
R.~Mesleh, H.~Elgala, and H.~Haas, ``On the performance of different {OFDM}
  based optical wireless communication systems,'' \emph{IEEE J. Opt. Commun.
  Netw.}, vol.~3, no.~8, pp. 620--628, August 2011.

\end{thebibliography}

\end{document}